\documentclass[journal,comsoc]{IEEEtran}
\usepackage[pdftex]{graphicx}
\usepackage{booktabs}
\usepackage[utf8]{inputenc}
\usepackage{amsmath}
\usepackage{amssymb}
\usepackage{subfigure}
\usepackage{array}
\usepackage{multirow}
\usepackage{float}
\IEEEoverridecommandlockouts

\begin{document}
\title{Integrating Sensing and Communications for Ubiquitous IoT: Applications, Trends and Challenges}

\author{Yuanhao~Cui,~\IEEEmembership{Member,~IEEE,}
Fan~Liu,~\IEEEmembership{Member,~IEEE,}
Xiaojun~Jing,~\IEEEmembership{Member,~IEEE,}\\
Junsheng~Mu,~\IEEEmembership{Member,~IEEE}}



\IEEEtitleabstractindextext{%
\begin{abstract}
Recent advances in wireless communication and solid-state circuits together with the enormous demands of sensing ability have given rise to a new enabling technology, integrated (radar) sensing and communications (ISAC). The ISAC captures two main advantages over dedicated sensing and communication functionalities: 1) \textit{Integration gain} to efficiently utilize congested wireless resources, and even, 2) \textit{Coordination gain} to balance dual-functional performance or/and perform mutual assistance. Triggered by ISAC, we are also witnessing a paradigm shift in the ubiquitous IoT architecture, in which the sensing and communication layers are tending to partially converge into a new layer, namely, \textit{the signaling layer}. 

In this paper, we first attempt to introduce a definition of ISAC, analyze the various influencing forces, and present several novel use cases. Then, we complement the understanding of the signaling layer by presenting several key benefits in the IoT era. We classify existing dominant ISAC solutions based on the layers in which integration is applied. Finally, several challenges and opportunities are discussed. We hope that this overview article will serve as a primary starting point for new researchers and offer a bird's-eye view of the existing ISAC-related advances.
\end{abstract}}
\maketitle
\IEEEdisplaynontitleabstractindextext
\IEEEpeerreviewmaketitle

\section{Introduction}
When the concept '\textit{Internet of Things (IoT)}' first emerged, its additional sensing capabilities were identified as a critical paradigm shift from computer networks \cite{kevin}. From then on, sensing and communications (S\&C), these two fundamental functionalities have been recognized to be indispensable in the design and implementation of ubiquitous IoT devices, ranging from autonomous vehicles, wearable electronics, and Wi-Fi to drones and satellites. In the current hardened-into-fixed IoT data processing pipelines, S\&C are individually accomplished by black-box-like modules, which do not necessarily share any external knowledge of their internal workings. This modularized IoT architecture encourages S\&C driving on two parallel layers (i.e., the sensing layer and the communications layer) with limited hardware intersection, little mutual assistance, and, therefore, rare integration.

Meanwhile, an unprecedented proliferation of new IoT services, e.g. extended reality (XR), digital twins, autonomous systems and flying vehicles, expresses a huge desire for novel sensing solutions. Wireless sensing capability enabled by analyzing received RF signal patterns and characteristics has the potential to become an essential component of the sensing solution. On the other hand, the combined use of high frequencies and large antenna array results in striking similarities between communication and radio sensing systems, in terms of the hardware architecture, channel characteristics, and information processing pipeline. Consequently, sensing and communications systems can be jointly designed, optimized, and dispatched to assist in each other or transmitted via the same hardware platform, common spectrum, joint signal processing strategy, and unified control framework.

With the various influences exerted from the technical and commercial perspectives, we anticipate that the sensing layer and communications layer are changing from separation to integration, which results in a paradigm shift in the IoT architecture. As a consequence, a new \textit{signaling layer} enabled by ISAC is emerging, with the advantages of low hardware cost, power consumption, and signaling latency as well as a small product size and improved spectral efficiency. Moreover, ISAC technology can provably endow current communications infrastructures with sensing functionalities while requiring minimal standard modifications, allowing existing communication networks to provide sensing and surveillance services for civilians. As a result, many new use cases can be made available in the contexts of autonomous vehicles, smart cities, smart homes, and cellular networks for 5G and beyond.

In this paper, we attempt to contribute to the concept of ISAC and to complement the understanding of the role of the signaling layer in the IoT architecture. We start by introducing our definition and understanding of ISAC by presenting fundamental principles and several key benefits. Then, we analyze the various forces influencing ISAC, followed by many novel use cases. In addition, we categorize and classify existing dominant ISAC solutions based on the layers in which the integration is taking place. Throughout the above, we also investigate and highlight related innovations and landmark advances reported from academia, industry, and standards associations in fields ranging from solid-state circuitry, microwave theory and techniques, signal processing, and communications to mobile computing. Finally, we overview and enumerate the practical challenges and open questions in this area. 


\begin{figure*}[bt]
\begin{minipage}[t]{1.0\linewidth}
		\centering
\centerline{\includegraphics[width = 0.95\linewidth]{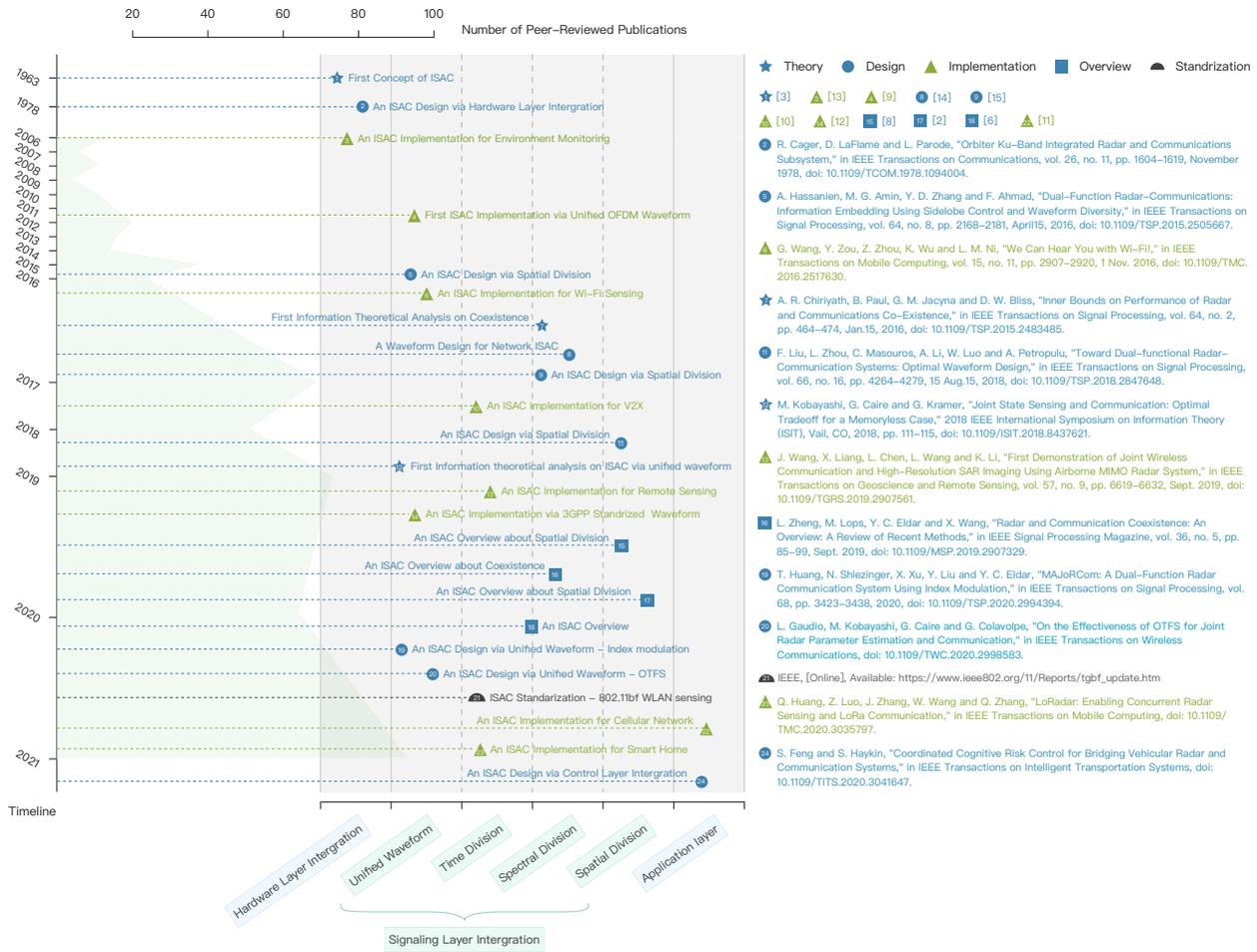}}
\end{minipage}
\caption{An illustration of ISAC-related research activities. The number of peer-reviewed publications is shown on the left-hand side. Several notable activities in contexts such as theory, design, implementation, overviews and standardization are indicated on the right-hand side. We also divide these landmarks in accordance with the categories given in Section V. The publication data was collected from IEEE Xplore in Jan. 2021.}
	\label{fig:spectrum_rad_comm}
\end{figure*}

\section{Why Integrating S\&C?}
An accelerating growth of research interest in ISAC is witnessed as shown on the left-hand side of Fig. 1.

\subsection{ What is ISAC? }

\textbf{ISAC} refers to a design paradigm in which (radio) sensing and communication systems are integrated to efficiently utilize congested resources and even to pursue mutual benefits, as well as the corresponding enabling technologies for this paradigm. We define ``integration'' as any combined use of two or more systems in whole or in part. For example, a wireless sensor network relies on hardware integration between sensing modules and communication modules in a distributed manner, a secondary surveillance radar system involves signaling integration between a surveillance radar and a communication transponder, and cognitive radar operating in the communication band requires spectrum integration. Although ISAC can improve the hardware, spectral, temporal, signaling, and energy efficiency of systems, the specific aspects in which integration is applied determine which resources can be saved.

\textbf{Levels of Integration.} The rationale of the ISAC is that a radio emission can simultaneously convey information from the transmitter to the receiver and extract information from the scattered echoes. Therefore, the unified communication and sensing waveform is considered to be the most tightly integrated configuration, in which all types of efficiency improvements can be achieved. Still, depending on the level at which the integration is taking place, as shown in Fig.~1, there are various benefits to be gained, including improved size-, hardware-, spectral-, energy-efficiency, and lower latency/signaling cost. Such looser configurations have also drawn numerous attentions from both industry and academia.


\textbf{Some Terminology.} Several terms have been used to describe the related research output, such as joint radar communications/joint communications radar (JRC/JCR) \cite{visa}, joint communication and radar/radio sensing (JCAS) \cite{zhang}, dual-functional radar communications (DFRC) \cite{hassanien}, radio-frequency (RF) convergence, and radar-communication (RadCom) \cite{sturm}. From our perspective, the aim of DFRC is to design novel waveforms to distinguish radar and communication functionalities via spatial division. RF convergence refers to broader radio integration that includes positioning, navigation, and timing systems. RadCom mainly focuses on endowing radar equipment with communication capabilities. JCAS is more concerned with the incorporation of sensing functionalities into the infrastructure side, particularly in cellular networks. In a sense, ISAC is interchangeable with the first two, but it focuses more on the paradigm shift of the network architecture and the effects on the electronics, the involved objects,
or the user-equipment side. Moreover, ISAC also emphasizes resource efficiency, while the others do not. 

\subsection{Why Do We Need ISAC in the IoT?}

\textit{1) Influence of Technical Trends}: Although the emergence of the ISAC concept can be traced back to 1960s \cite{Mealey}, when coded pulses were employed to convey information from a ground radar to a space vehicle, there was a paucity of further developments in the following decades. We tend to attribute this observed stagnation to the use of dedicatedly designed RF circuits at the time, meaning that previous RF devices tended to be specific to the domain of either radio (radar) sensing or communications.

\textbf{Hardware.} With recent advances in solid-state circuits and microwave technology, however, the hardware feasibility of leveraging radio sensing in tiny IoT products tends to no longer be a barrier. For example, a multiple-input multiple-output (MIMO) radar system-on-a-chip (SoC) constructed from 192 virtual receivers
has been reported to achieve a $\pm 1^{\circ}$ angular resolution and a $0.099$ km/h Doppler resolution, within silicon areas of only 14 $\text{mm}^2$ for 12 mmWave transceivers and 71 $\text{mm}^2$ for the overall SoC \cite{Giannini}. These key performance indicators already meet the requirements for various radio sensing use cases, as shown in
Fig.~2. Thus, it is safe to infer that the integration of S\&C circuits at the chip level, i.e., ISAC SoCs for mobile devices or ISAC baseband processors, will emerge in the next few years.

\textbf{Signal Processing.} The combined use of mmWave frequencies and massive MIMO technology results in striking similarities between communication and radio sensing systems in terms of the hardware architecture\footnote{To reduce the cost of RF chains in MIMO configurations, existing commercial mmWave RF front-ends are implemented with phase shifters together with variable gain amplifiers. This has led to the emergence of phased-MIMO radar and hybrid beamforming techniques in the radar and communications literature, respectively.}, channel characteristics, and information processing pipeline. Moreover, with the development of mmWave
technology and beam-domain signal processing strategies, it has become possible to straightforwardly extend several radar missions, e.g., angle of arrival (AoA) estimation, angle of departure (AoD) estimation, and moving target tracking, to address emerging communication challenges, e.g., beam management \cite{fanliu}. It is reasonable to envision that the reuse of signaling strategies between the S\&C functionalities can lead to mutual benefits.

\textbf{Mobile Computing.} Current Wi-Fi sensing applications require the extraction of multipath channel information from the raw CSI measurements, since this multipath information is the principal component that captures how the surrounding environment changes. In general, the raw CSI measurements have been compensated in Wi-Fi baseband processors, e.g., by means of the sampling time offset (STO), to synchronize the oscillator clocks of the transmitter and receiver. However, such offsets are hidden
in a communications black box and are thereby unknown to the sensing modules. Consequently, time and frequency offsets create ambiguity when a sensing module calculates range/velocity estimates and increase the false alarm probability
when recognizing human activities. To address this issue, an additional processing procedure of fitting and then removing the clock/frequency offsets is employed. However, these offsets can instead be straightforwardly removed by breaking the cross-system isolation and exchanging the necessary information between the S\&C functionalities in the baseband processor.


\begin{figure*}[ht!]
\begin{minipage}[t]{1.0\linewidth}
		\centering
\centerline{\includegraphics[width = 0.9\linewidth]{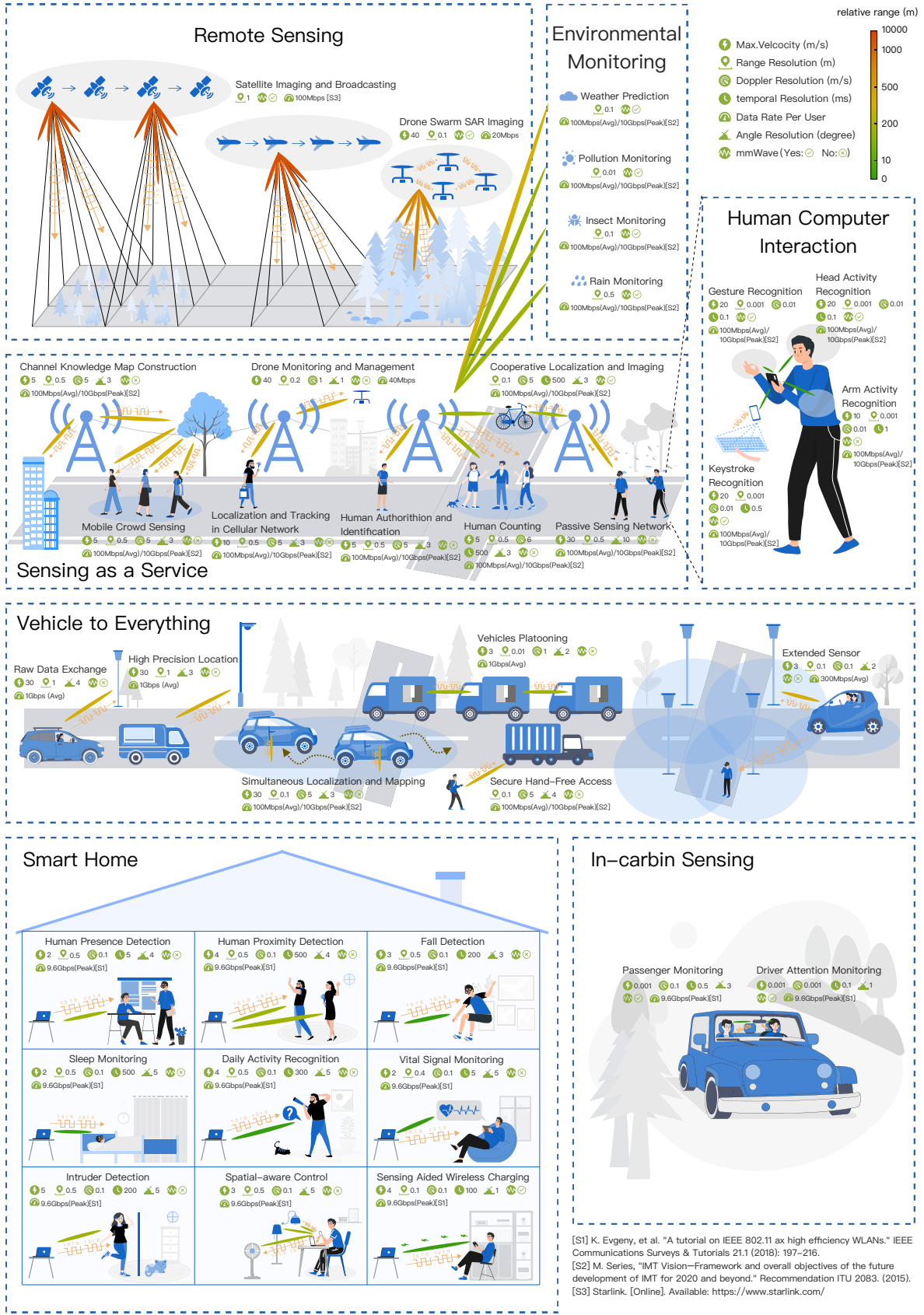}}
\end{minipage}
\caption{Seven scenarios and 34 use cases of ISAC. The required key parameter indicators, such as the maximum velocity, range/Doppler/temporal/angular resolutions, and data rate resolution, are marked in the legends. The beam colors indicate the maximum ranges in the various use cases.}
	\label{fig:usecases}
\end{figure*}

\textit{2) Influence of Commercial and Regulatory Forces}: Novel ISAC applications have already gone far beyond academic studies, particularly in regard to Wi-Fi sensing applications. 

\textbf{Commercial Progress.} The CSI measured in Wi-Fi networks has been widely analyzed to support various short-range sensing tasks in a device-free\footnote{Here, `device-free' means without reliance a connection with user's devices or requiring them to carry any device.} manner. For example, Wi-Fi devices can detect the presence of humans in a conference room (with an accuracy of $97\%-100\%$); recognize human activities (with an accuracy of $73\%-100\%$, depending on the set of activities considered) such as walking, running, and exercising; and even imaging surrounding objects (with an imaging error of $<4.5~cm/\pm 1^{\circ}$). Furthermore, according to Intel, WLAN sensing is recognized as a key direction of development toward Wi-Fi 7.

\begin{table*}[]
\centering
\scriptsize
\caption{Descriptions of Use Cases and Challenges}
\label{tab:my-table}
\resizebox{\textwidth}{!}{%
\begin{tabular}{p{0.08\linewidth}|p{0.2\linewidth}|p{0.45\linewidth}|p{0.15\linewidth}}
\hline
Scenario & Use Case & Descriptions & Case Challenges \\ \hline
\multirow{9}{\linewidth}{Smart Home} & Human  Presence Detection & \multirow{9}{\linewidth}{\begin{tabular}[c]{@{}p\linewidth@{}}Amplitude/phase variations of wireless signal   could be employed to detect or recognize human   presence/proximity/fall/sleep/breathing/daily activities, by extracting the   range, Doppler, or micro-Doppler features while moving indoor.\\      \\       Spatial position sensed by wireless signals could be employed as a piece of prior knowledge to assist in   location-aware control or wireless powered communication.\end{tabular}} & \multirow{9}{\linewidth}{\begin{tabular}[c]{@{}p\linewidth@{}}Low range/time resolution of standard   waveform;\\       Fixed pilot allocation;  \\      Clock Synchronization;\\      Narrow band signal yields low Doppler resolution;\\      Transitory behavior;\\      Assistance from other sensors, e.g. camera.\end{tabular}} \\ \cline{2-2}
 & Human Proximity Detection &  &  \\ \cline{2-2}
 & Fall Detection &  &  \\ \cline{2-2}
 & Sleep Monitoring &  &  \\ \cline{2-2}
 & Daily Activity Recognition &  &  \\ \cline{2-2}
 & Breathing/Heart Rate Estimation &  &  \\ \cline{2-2}
 & Intruder   Detection &  &  \\ \cline{2-2}
 & Location-aware Control &  &  \\ \cline{2-2}
 & Sensing Aided Wireless Charging &  &  \\ \hline
\multirow{8}{\linewidth}{Sensing as a Service \cite{zhang}} & Drone   Monitoring and Management & \multirow{8}{\linewidth}{\begin{tabular}[c]{@{}p\linewidth@{}}Empowered by the ISAC, IoT devices, and cellular   network are able to provide sensing services to civilians, including enhanced   network localization, cooperative imaging in a given area, and mobile crowd   sensing, etc.\\     Human counting, authoritarian, and identification services could also be   constructed by recognizing the influence patterns to the channel state   information, which is specific to a certain scenario.\\      Densely deployed ISAC transceivers can naturally form a passive sensing network to assist in nearby devices.\\     Each base station is able to measure the surrounding environment to   construct a site-specified channel knowledge database, to speed up beam   alignment procedure.\end{tabular}} & \multirow{8}{\linewidth}{\begin{tabular}[c]{@{}p\linewidth@{}}Low  range/time resolution of standard waveform;\\      Wireless resource allocation and optimization;\\      Interference management;\\      Cooperative sensing and imaging design;\\      Small-Size drone recognition and tracking.\end{tabular}} \\ \cline{2-2}
 & Localization and Tracking in Cellular Network &  &  \\ \cline{2-2}
 & Human Authorization and Identification &  &  \\ \cline{2-2}
 & Human Counting &  &  \\ \cline{2-2}
 & Area Imaging &  &  \\ \cline{2-2}
 & Mobile Crowd Sensing &  &  \\ \cline{2-2}
 & Channel Knowledge Map Construction &  &  \\ \cline{2-2}
 & Passive Sensing Network &  &  \\ \hline
\multirow{4}{\linewidth}{Human Computer Interaction} & Gesture   Recognition & \multirow{4}{\linewidth}{The object’s characteristics and dynamics can be captured from the time/frequency/Doppler variations of the reflected signal. Therefore, detect touchless gesture interactions via wireless signal is a new human-computer interaction technology.} & \multirow{4}{\linewidth}{\begin{tabular}[c]{@{}p\linewidth@{}}Transitory   behavior;\\      High frame rate requirement;\\      Beam width optimization.\end{tabular}} \\ \cline{2-2}
 & Keystroke Recognition &  &  \\ \cline{2-2}
 & Head Activity Recognition &  &  \\ \cline{2-2}
 & Arm Activity Recognition &  &  \\ \hline
\multirow{6}{\linewidth}{\begin{tabular}[c]{@{}p\linewidth@{}}Vehicle to Everything\\      (V2X)\cite{kumari} \end{tabular}} & Raw   Data Exchange and High Precision Localization & \multirow{6}{\linewidth}{\begin{tabular}[c]{@{}p\linewidth@{}}ISAC aided V2X could simultaneously perform high-rate communications and high-precision localization.\\         ISAC aided V2V communication provide environmental information to support fast vehicle platooning, secure access,   simultaneous localization and mapping. \\      Roadside units (RSUs) network can provide sensing  services to extend the sensing range of a passing vehicle beyond its own line   of sight and field of view.\end{tabular}} & \multirow{6}{\linewidth}{\begin{tabular}[c]{@{}p\linewidth@{}}Full-duplex   problem;\\      Protocol design for vehicle communications;\\     Multi-source sensing information fusion;\\      Sensing aided vehicular communications.\end{tabular}} \\ \cline{2-2}
 & Secure Hand-Free Access &  &  \\ \cline{2-2}
 & Vehicle Platooning &  &  \\ \cline{2-2}
 & Simultaneous Localization and Mapping &  &  \\ \cline{2-2}
 & Extended Sensor &  &  \\ \hline
\multirow{3}{\linewidth}{In-cabin Sensing} & Passenger   Monitoring & \multirow{3}{\linewidth}{The   micro motion of faces could 'modulate' the    wireless signals, and then, some RF features such as CSI measurements   are extracted to analyze the attention and activities of drivers and   passengers.} & \multirow{3}{\linewidth}{\begin{tabular}[c]{@{}p\linewidth@{}}Noise   reduction;\\      Transitory behavior;\\      High frame rate.\end{tabular}} \\ \cline{2-2}
 & Driver Attention &  &  \\ &Monitoring &  &  \\ \hline
\multirow{3}{\linewidth}{Remote Sensing} & Drone   Swarm SAR Imaging & \multirow{3}{\linewidth}{A swarm of drones can cooperate to act as a mobile antenna array. Synthetic aperture imaging can be performed to achieve high-resolution all-weather  day-and-night imaging.} & \multirow{3}{\linewidth}{\begin{tabular}[c]{@{}p\linewidth@{}}  Trajectory 
optimization;\\      Joint S\&C control; \\      Resource allocation.\end{tabular}} \\ \cline{2-2}
 & Satellite Imaging &  &  \\ & and Broadcasting &  &  \\ \hline
\multirow{4}{\linewidth}{Environmental Monitoring \cite{science}} & Weather   Prediction & \multirow{4}{\linewidth}{\begin{tabular}[c]{@{}p\linewidth@{}}By analyzing the path loss of mmWave links,   variations in environmental characteristics such as water vapor, air   pollutants, and insects can be observed. \\       A cellular network with a sensing function can serve as a built-in   monitoring facility and can be utilized as a 
large-scale atmospheric observation network.\end{tabular}} & \multirow{4}{\linewidth}{\begin{tabular}[c]{@{}p\linewidth@{}}Large-scale data analysis;\\      Distributed beamforming and optimization.\end{tabular}} \\ \cline{2-2}
 & Pollution Monitoring &  &  \\ \cline{2-2}
 & Rain Monitoring &  &  \\ \cline{2-2}
 & Insect Monitoring &  &  \\ \hline
\end{tabular}%
}
\end{table*}

\textbf{Spectrum Regulatory Aspect.} Another strong force driving ISAC forward is exerted by the vast commercial requirements on radio sensing. Unfortunately, novel civilian radio sensors bear a disproportionate regulatory burden. For instance, the Federal Communications Commission (FCC) granted the spectrum allocation request of the Soli project only after a year-long discussion, and at present, Soli is still not allowed to operate in many major countries, such as Japan, India, and China. Moreover, radio sensing and communications functionalities in large IoT devices tend to operate in shared and often congested or even contested spectra, e.g., 5G-based IoT devices vs. military radar in the 3.5 GHz band and mmWave automotive radar vs. mmWave 5G communication in the 60 GHz band. To help overcome these conflicts, ISAC
can conveniently enable communication devices to sense the environment while sharing the same spectrum \cite{zhang}.


\textit{3) A Number of Use Cases}: ISAC-enabled IoT devices are expected to promote many new applications. We illustrate seven scenarios and 34 use cases in Fig.~2, where their key parameter indicators are also marked. Their descriptions and challenges are shown in Table.~1.

\begin{figure*}[ht!]
\begin{minipage}[t]{1.0\linewidth}
		\centering
\centerline{\includegraphics[width = 0.9\linewidth]{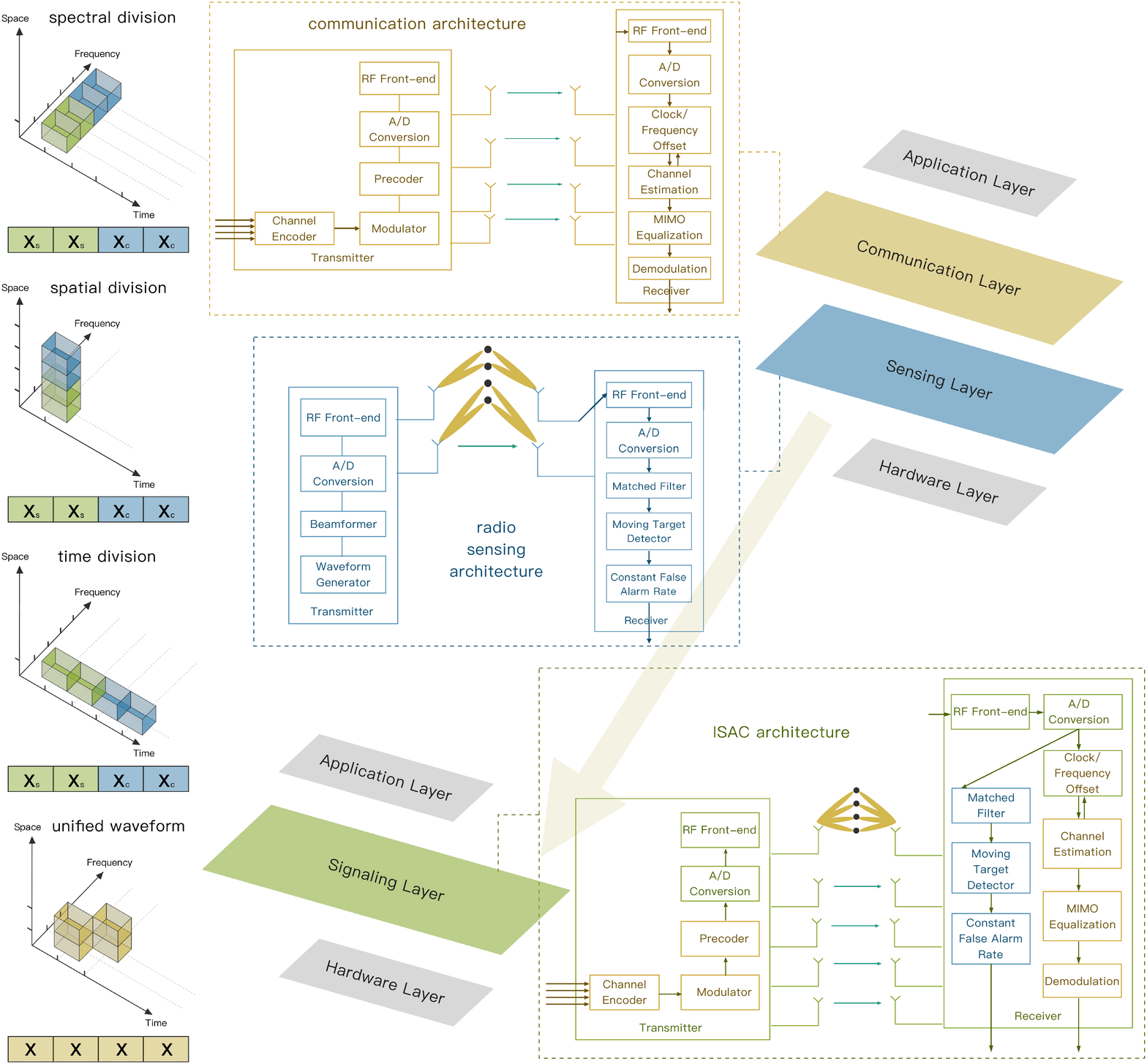}}
\end{minipage}
\caption{To illustrate the paradigm shift driven by ISAC, we mark the communication-specific architecture in yellow, the sensing-specific architecture in blue, and the architecture shared between S\&C in green. The left-hand side of this figure shows the four signaling strategies discussed in Section V-B. $\mathbf{X}_s$ and $\mathbf{X}_c$ denote the resources allocated to the S\&C functions, respectively.}
	\label{fig:dof}
\end{figure*}


\section{Leveraging ISAC in the IoT: A Paradigm Shift}

In a conventional IoT information processing pipeline, environmental information is collected by sensors, exchanged via communications, and fused by processing units to support environment-aware decision-making and intelligent human-computer interaction. Consequently, a generic IoT architecture consists of three layers: 1) a sensing (perception) layer, to collect, process, and digitize environmental information to obtain voluminous data; 2) a communication (transport) layer, to convey the sensing data to the network or the application layer; and 3) an application layer, to employ computing techniques to extract valuable information collected by the current device itself or transmitted from other devices. Data mining and machine learning are usually applied in the application layer to achieve autonomy of IoT devices. Traditionally, radio sensing belongs to the sensing layer, and wireless communication occurs in the communication layer. This communication-after-sensing design philosophy fetters the combined usage of S\&C, particularly for information sharing and co-designed signaling strategies.

Due to their ISAC-related advantages, joint signaling strategies are provably able to overcome interlayer constraints and enable optimal co-design and operation \cite{visa}. As such strategies represent the most tightly integrated setting, it is reasonable to envision that the sensing and communication layers are currently tending to partially converge toward signaling-level integration, particularly between wireless sensing and wireless communication, via waveform unification. 

\textbf{Signaling Layer}. The resulting operation layer is named the \textit{signaling }layer, as shown in Fig.~3. Compared to the conventional IoT architecture, the new signaling layer is intended to handle radio emission and related post-processing signaling strategies, including S\&C functionalities, and thereby perform the information extraction and data transmission tasks as well as was possible originally. Joint signal processing for both S\&C should be conducted in this layer to allow full access to all necessary information.


Benefiting from the abandonment of interlayer isolation, the signaling layer permits the efficient exchange of useful information between the S\&C functions. Moreover, without the constraints imposed by dedicated functionalities, more design degrees of freedom (DoFs) can be accessed via co-design to flexibly optimize operation parameters, balance resource allocation, and even mutually assist in improving the capabilities of the radio sensing and communication functionalities.  

In particular, Wi-Fi devices equipped with an antenna array could achieve fine-tuning of the beamwidth and beam direction to form a sharp, pencil-like sensing beam focusing only on a certain
object of interest. Such a highly directional beam could lessen the floor/wall reflections affecting human activity recognition, thereby allowing the sensing signal to convey pure activity information to the receiver side. Consequently, compared to the original black-box-like sensing and communication layers, the new signaling layer can serve as a more adaptable and robust 
backbone to support high-layer applications.

\section{Features of ISAC in the IoT}

Although a number of studies to date have offered descriptions of ISAC-related designs and applications, few attempts have been made to systematically quantify the advantages of ISAC. Here, we briefly discuss the main potential advantages of ISAC in the IoT era.


\subsection{Integration Gain}
Integration gain is the fundamental reason for the superiority of ISAC over separate S\&C functionalities, especially for IoT devices. In essence, the integrated operation in ISAC means that the components or resources used for the S\&C functionalities can be coupled to achieve more efficient resource utilization. Depending on the level at which integration is applied, there are various benefits to be gained, including improved size, hardware, spectral, and energy efficiency as well as lower latency and signaling costs.

For example, ISAC can be achieved via signaling layer integration while splitting the antenna array and RF chains into two groups: one for radar and one for communication. Compared to a shared antenna array, i.e., a more tightly coupled ISAC setting, the reduced spatial DoFs, lower angular resolution, and additional interference management constraints impose extra expenses for both S\&C \cite{fanliu2}. However, when we compare this separated antenna setting with the case of entirely separate sensing and communication layers, i.e., a more loosely coupled setting, it shows improved energy and hardware efficiency.

Spectral efficiency can be readily pursued by means of various spectrum sharing approaches, e.g., cognitive radio. Indeed, interference management \cite{visa} using a joint signaling strategy is precisely the signaling layer integration approach for spectrum integration. Moreover, signaling strategies can be coupled more tightly to allow S\&C to be performed simultaneously on the basis of a single radio emission. In this way, a dual-mission signal can be prepared in which the two functionalities are allocated over non-overlapped resources \cite{kumari,hassanien}, or are even achieved with a fully unified waveform \cite{sturm}.

\subsection{Coordination gain}
Both communication and radio sensing require the acquisition of situational awareness concerning the surrounding radio environment and equipment, commonly in the form of CSI, the directions of the desired users, targets or interferers, and even a map of the surrounding channel knowledge. Depending on the level of integration, various types of information can be conveniently shared in a cross-function or cross-user manner, e.g., in a shared memory or in the same processor, to jointly design signal processing strategies for balancing performance or achieving mutual assistance. Benefit from this, ISAC practically permits full exploitation of the design DoFs, while the performance of other waveforms is constrained by, for example, standards.


For example, conventional RSUs are equipped with mmWave massive MIMO vehicle-to-everything (V2X) communication systems and sensors, e.g., cameras or radar, and are willing to serve passing high-mobility vehicles with reliable and large-capacity
data transmission. Due to the high directionality of narrow ``pencil-like'' mmWave beams, beam misalignment can readily occur and severely compromise the communication transmission rate \cite{heath}. If a frequency-independent representation of the spatial signal paths from sensor to vehicle\footnote{
A frequency-independent representation of signal paths can usually be obtained by estimating spatial-related channel parameters, e.g., AoA, path attenuation, frequency-independent phase shift, and distance.} can be obtained, this situational information can be reused by the communication system to improve beam tracking performance, even though the S\&C systems are only loosely coupled in such a case.

Things become more interesting when the S\&C functionalities are coupled more tightly, i.e., the knowledge of the transmitted communication symbols is synchronized with the sensing function. In essence, such a shared communication signal plays the role of prior knowledge in the sensor's post-processing procedure, i.e., a reference signal, which could be employed to construct a partially matched filter and perform pulse compression to improve the sensing function's detection probability. Additionally, when a ISAC transceiver emits unified waveform to a communication receiver, the CSI information may be estimated and inferred from sensing echoes, resulting in a pilot-free signaling strategy and thus making it possible to take advantage of lower latency and signaling costs\cite{fanliu}.

\section{Dominant ISAC Solutions}

In this section, we attempt to bridge the new IoT architecture paradigm discussed above with the current dominant ISAC solutions.


\subsection{Hardware Layer Integration}
\textbf{Software Defined Radio}. Running different transmitter/receiver strategies on the same device is not new. Thanks to the development of software-defined radio (SDR) in the mid-2000s, signal processing can now be handed over from special-purpose RF circuits to general-purpose processors. From then on, hardware reuse for S\&C can be accomplished by means of SDR for large IoT devices such as autonomous vehicles. In such a case, the radio system dynamically generates complex signals adhering to different standards or even signals that are not standardized. However, reconfiguration operations for SDR are typically time consuming and independent of any particular function. Thus, the lack of co-designed hardware/signaling strategies leads to rare coordination gain.


 \textbf{ISAC System-on-chip/in-package}. Multichannel RF transceivers and high-performance analog-to-digital converters (ADC) could be integrated as a radar or communications SoC. Moreover, antenna array has been realized by a radar or communications system-in-package (SiP) solution in several tiny IoT devices. It is safe to envision that S\&C functionalities would be integrated in a chipset via SoC/SiP, to pursue high integration gain as well as coordination gain.


\subsection{Signaling Layer Integration}

When S\&C functions share the same spectrum, the communication propagation characteristics are much akin to those of radio sensing. Thus, although they have drastically different purposes, the existing signaling strategies for S\&C show several common features, especially in terms of waveform and beamforming designs. Inspired by multiaccess technology, it is natural to consider that these two functionalities can be harmonized into a single emission via orthogonal/non-overlapped resource allocation, e.g., time/frequency/spatial division. As a step further, fully unified waveform tends to be a more favorable design, which is able to more efficiently utilize wireless resources and thereby improve the integration gain. Below we elaborate on these two aspects.

\subsubsection{Non-overlapped Resource Allocation}

\begin{itemize}
\item{\textbf{Time Division}}: The most straightforward ISAC approach is to schedule S\&C waveforms in different time slots, where the two functionalities are loosely coupled. Indeed, the time-division ISAC solution has been widely employed to add sensing capabilities to existing communication protocols, such as the 802.11p and 802.11ad standards \cite{heath}, and has shown great potential to empower cellular IoT devices with sensing functions in a fast and inexpensive manner \cite{zhang}. Notably, most Wi-Fi sensing approaches rely on pilot signals, which are transmitted in a time-division fashion in conjunction with payload data.
\item{\textbf{Spectral Division}}: An alternative strategy is to allocate the S\&C waveforms to different subcarriers or different frequency bands. A subcarrier selection indicator with elements taking values of 0 or 1 can be employed to map the sensing waveform and communication waveform to different subcarriers. Typically, frequency-division-based ISAC solutions can be employed in existing commercial orthogonal frequency-division multiplexing (OFDM) systems with only minor modifications. 
\item{\textbf{Spatial Division}}: Another widely investigated signaling strategy is to form multiple spatial beams for simultaneously serving communication receivers and performing sensing tasks such as target detection. In general, spatial division can be realized by selecting a sensing waveform that lies in the other system's null space. However, the null space of the channel is determined by the radio propagation environment and cannot be controlled by the designer. Therefore, interference CSI (ICSI) estimation is essential for both waveform and spatial filter design. 
\end{itemize}

\subsubsection{Fully Unified Waveform}
As the most tightly integrated setting, the design of a fully unified ISAC waveform with the shared use of wireless resources is the most desirable case, as it offers the potential for the highest integration and coordination gains. In general, the unified waveform can be designed following three philosophies, namely, sensing-centric design, communication-centric design, and joint design.
\begin{itemize}
\item{\textbf{Sensing-Centric Design:}} In the event where the primary function is sensing, e.g., to equip a radar sensor with the communication ability, ISAC can be implemented on existing sensing waveforms or infrastructures, which requires to embed communication data into a radar waveform over different signal domains. A classical ISAC waveform design is to modulate communication symbols onto chirp carriers, where ASK/PSK/FSK data can be embedded in the time-frequency domain \cite{fanliu}. Furthermore, direct spread spectrum sequences (DSSS) for CDMA communications can be naturally combined with phase-coded radar waveforms to produce code-domain ISAC waveforms \cite{fanliu}. More recently, spatial domain is exploited for ISAC, where the useful information can be represented by the sidelobe level of a radar beampattern, or by the antenna indices of a MIMO radar \cite{hassanien}. Nonetheless, sensing-centric design can only be applied to limited communication scenarios, as it suffers from low transmission rate, which is tied to the pulse repetition frequency (PRF) of the radar. 
\item{\textbf{Communication-Centric Design:}} Communciation-centric design refers to exploiting existing communication waveform directly for radar sensing. In principle, any communication waveform can be leveraged for mono-static ISAC signaling, as the transmitted data and waveform are known {\emph{a priori}} at the transmitter. A pioneering communication-centric design is to employ OFDM communication waveform for target detection, where the range and Doppler parameters can be readily obtained via IFFT and FFT \cite{sturm}. Since the communication waveform is not tailored for radar, its sensing performance is rather limited, where sophisticated signal processing techniques are needed to compensate for the performance loss. Moreover, communication waveform needs to be carefully shaped to satisfy specific sensing constraints, e.g., low PAPR, good correlation properties, and reliance to clutter interference, etc.
\item{\textbf{Joint Design:}} Instead of relying on existing S\&C waveforms, one may also conceive a joint ISAC waveform from the ground-up, such that a flexible performance trade-off can be readily achieved. This method is known as joint design, for which the sensing-centric and communication-centric designs can be viewed as two extreme cases. Joint ISAC waveform design can be formulated as a mathematical optimization problem, where the objective function is sensing/communication performance metrics, with the constraints to guarantee the performance of the other functionality. As a example, in \cite{fanliu2}, the MIMO radar beampattern is optimized, subject to per-user SINR constraints for communications.
\end{itemize}


\begin{table*}[]
\centering
\caption{An Overview of Open Challenges in ISAC}
\label{tab:my-table2}
\resizebox{\textwidth}{!}{%
\begin{tabular}{|c|c|c|c|}
\hline
\multicolumn{3}{|c|}{Layers} & Open Problems \\ \hline
\multicolumn{3}{|c|}{Hardware Layer} & \begin{tabular}[c]{@{}c@{}}Dual-functional RF Front-End\\  Full-Duplex Receiver\\ISAC System-on-chip/in-package\\ Low-Complexity Hybrid Analog-Digital Structure\end{tabular} \\ \hline
\multirow{6}{*}{Signaling Layer} & \multirow{3}{*}{Non-overlapped Resource Allocation} & Time Division & S\&C Event Scheduling and Resource Allocation \\ \cline{3-4} 
 &  & Spectral Division & \begin{tabular}[c]{@{}c@{}}Subcarrier Assignment and Allocation\\ Coded Waveform\end{tabular} \\ \cline{3-4} 
 &  & Spatial Division & \begin{tabular}[c]{@{}c@{}}Joint Transmit-Receive Beamforming\\ Low-Complexity Cooperative Precoding\end{tabular} \\ \cline{2-4} 
 & \multirow{3}{*}{Fully Unified Waveform} & Sensing-Centric Design & \begin{tabular}[c]{@{}c@{}}Cooperative Sensing Waveform\\ Information Embedded Waveform\\ Interference Management in ISAC Network\end{tabular} \\ \cline{3-4} 
 &  & Communication-Centric Design & \begin{tabular}[c]{@{}c@{}}ISAC Frame Structure\\  ISAC Protocol Design\\  ISAC Transceiver Solutions\\  Networked ISAC\end{tabular} \\ \cline{3-4} 
 &  & Joint Design & \begin{tabular}[c]{@{}c@{}}Fundamental Limits and Tradeoff between S\&C\\ Unified Evaluation Metric\\ Practical ISAC Parameter Adjustment \\ Joint Signal Processing Strategies\\  Performance Analysis of ISAC\end{tabular} \\ \hline
\multicolumn{3}{|c|}{Application Layer} & \begin{tabular}[c]{@{}c@{}}RF Feature Extraction\\ Human Activity Recognition\\ Joint S\&C Control Circle\\ RF Imaging Algorithm\\ Simultaneous Communication, Localization, and Mapping\\ Reflective Intelligent Surfaces (RIS)-aided ISAC\end{tabular} \\ \hline
\end{tabular}%
}
\end{table*}
\subsection{Application Layer Integration}

Combining the S\&C functions in the application layer mainly allows each function to exploit the output data of the other to obtain insightful information for mutual assistance. In a typical sensing-after-communication or communication-after-sensing processing pipeline, such application layer integration usually operates in a serial fashion and thus is loosely coupled.

\textbf{Human Activity Recognition}. The widely investigated Wi-Fi sensing approach is a typical example of application layer integration. In current real-world Wi-Fi sensing methods, the raw CSI measurements are first transferred to an additional signal processing unit for data augmentation, e.g., noise reduction and outlier removal. Second, the target signal is extracted from the augmented CSI measurements by means of thresholding, filtering, or signal compression to remove redundant signals. Finally, model-based or learning-based algorithms are used to analyze the cleansed data to perform human activity detection or recognition.

\textbf{Coordinated S\&C Control}. In a S\&C systems that coordinated by a unified control center, one system can learn from and react to the risks that the other system has encountered. For example, benefiting from the coordinated cognitive risk control framework \cite{haykin}, the tracking accuracy of a vehicular radar system can be improved based on the situational information shared by a coordinated communication system, while in turn, the communication system can be made more efficient and reliable against attacks. 
Further coordination gain may be achieved by employing advanced data mining and cognitive-related techniques to control and optimize sensing or communication's performance, replying on the other's output.

\section{Open Challenges and Future Directions}

We overview recent ISAC open challenges in Table.~II, and then elaborate several of them in the following.

\textbf{Fundamental Limitations and Trade-offs of ISAC}: A theoretical performance analysis is critical for evaluating the superiority of current ISAC solutions. A unified upper bound and its achievability are now required to provide a general analysis framework and design objective for S\&C. One possible direction of investigation is to bridge information theory with detection theory. Due to the inherent relation between mutual information and the minimum mean square error (MMSE), it is expected that a closed-form expression can be derived to reveal the relationship between the performance of sensing estimation and the communication channel capacity, possibly in the form of a connection between the channel capacity and Cram\'er-Rao Lower Bound (CRLB).


\textbf{Practical ISAC Parameter Adjustment}: In practice, the signaling layer is the back-end that generates data for use in the application layer. It is reasonable to infer that better data quality leads to better application performance. For good sensing performance, the sensing beam should be fully focused on the target of interest to avoid redundant signal paths, e.g., reflection from the floor. From the communication perspective, however, more reflection paths are necessary to achieve a multiplexing gain. Thus, the balance between S\&C performance is subject to many practical constraints. For example, the parameters of the sensing function should be adjusted depending on the shape of the target while respecting communication quality concerns. In addition, the pulse repetition frequency determines the temporal resolution, i.e., the frame rate, which is of critical importance in several scenarios, such as gesture recognition.

\textbf{ISAC Receiver Solutions}: In an ISAC receiver, distinguishing sensing echoes from received communication signals is a challenging task, particularly in a rich scattering environment such as an indoor scenario. Most existing estimation technologies cannot be straightforwardly applied for this purpose. One possible solution is to allocate sensing echoes and uplink data to different time slots by employing a proper ISAC protocol.

\textbf{Networked ISAC}: Intuitively, multiple ISAC IoT devices can function as a multistatic radar to perform joint sensing of a target or a specified area. In this way, several sensing tasks, such as imaging, may be accomplished. For such cases, the information exchange and cooperative sensing processes between nodes have yet to be investigated. Moreover, the assignment and management of S\&C beams are also critical tasks in practice, particularly for ubiquitous IoT devices.


\section{Conclusion}

In this article, we have presented our understanding and definition of ISAC. To this end, we started by analyzing the forces driving the development of ISAC, followed by illustrations of many new use cases. It was shown that ISAC may lead the communication and sensing layers to partially converge into a new signaling layer. In this way, several advantages can be achieved, such as low hardware cost, power consumption, and signaling latency as well as a small product size and improved spectral efficiency. After that, we highlighted the gains that may be achieved with ISAC, including integration gains in terms of resource efficiency and coordination gains in terms of mutual assistance. We also discussed several major ISAC solutions, ranging from hardware layer integration and signaling layer integration to application layer integration. Finally, key challenges, opportunities, and directions of future research related to ISAC implementation were discussed, yielding the conclusion that ISAC will play an essential role in the IoT era.


%

\appendices



\ifCLASSOPTIONcaptionsoff
\newpage
\fi




\begin{thebibliography}{1}
\bibitem{kevin}
A. Kevin., “That ‘internet of things’ thing." RFID journal, vol. 22, no. 7, June 2009, pp. 97-114.
\bibitem{visa}
K. V. Mishra et al., “Toward Millimeter-Wave Joint Radar Communications: A Signal Processing Perspective,” in IEEE Signal Processing Mag., vol. 36, no. 5, Sept. 2019, pp. 100–14.
\bibitem{Mealey}
R. M. Mealey, "A Method for Calculating Error Probabilities in a Radar Communication System," in IEEE Transactions on Space Electronics and Telemetry, vol. 9, no. 2, June 1963, pp. 37-42.
\bibitem{Giannini}
V. Giannini et al., "A 192-Virtual-Receiver 77/79GHz GMSK Code-Domain MIMO Radar System-on-Chip," 2019 IEEE International Solid- State Circuits Conference - (ISSCC), San Francisco, CA, USA, 2019, pp. 164-166.
\bibitem{zhang}
A. Zhang et al., "Perceptive Mobile Network: Cellular Networks With Radio Vision via Joint Communication and Radar Sensing," in IEEE Vehicular Technology Magazine, doi: 10.1109/MVT.2020.3037430.
\bibitem{fanliu}
F. Liu et al., “Joint Radar and Communication Design: Applications, State-of-the-Art, and the Road Ahead,” in IEEE Transactions on Communications, vol. 68, no. 6, June 2020, pp. 3834–62. 
\bibitem{fanliu2}
F. Liu, et al., "MU-MIMO Communications With MIMO Radar: From Co-Existence to Joint Transmission," in IEEE Transactions on Wireless Communications, vol. 17, no. 4, pp. 2755-2770, April 2018.
\bibitem{hassanien}
A. Hassanien et al., “Dual-Function Radar Communication Systems: A Solution to the Spectrum Congestion Problem,” in IEEE Signal Processing Mag., vol. 36, no. 5, Sept. 2019, pp. 115–26.
\bibitem{sturm}
C. Sturm et al., “Waveform Design and Signal Processing Aspects for Fusion of Wireless Communications and Radar Sensing," in Proceedings of the IEEE, vol. 99, no. 7, July 2011, pp. 1236-1259.
\bibitem{kumari}
P. Kumari et al., "IEEE 802.11ad-Based Radar: An Approach to Joint Vehicular Communication-Radar System," in IEEE Transactions on Vehicle Technology, vol. 67, no. 4, April 2018, pp. 3012-3027.
\bibitem{heath}
A. Ali et al., "Leveraging Sensing at the Infrastructure for mmWave Communication," in IEEE Communications Magazine, vol. 58, no. 7, July 2020, pp. 84-89.
\bibitem{cui}
Y. Cui et al., "Interference Alignment Based Spectrum Sharing for MIMO Radar and Communication Systems," in the proceeding of 2018 IEEE 19th International Workshop on Signal Processing Advances in Wireless Communications (SPAWC), Kalamata, 2018, pp. 1-5.
\bibitem{science}
H. Messer et al., “Environmental Monitoring by Wireless Communication Networks,” Science, vol. 312, no. 5774, 2006, pp. 713–713.
\bibitem{beca}
M. Bica, et al., "Mutual Information based Radar Waveform Design for Joint Radar and Cellular Communication Systems," in  the proceeding of 2016 IEEE International Conference on Acoustics, Speech and Signal Processing (ICASSP), Shanghai, 2016, pp. 3671-3675.
\bibitem{haykin}
S. Feng, et al., "Coordinated Cognitive Risk Control for Bridging Vehicular Radar and Communication Systems," in IEEE Transactions on Intelligent Transportation Systems, doi: 10.1109/TITS.2020.3041647.
\end{thebibliography}
%

\end{document}